\documentclass[preprint,12pt,2023]{elsarticle}
\usepackage[utf8]{inputenc}
\usepackage{amsmath,bm}
\usepackage{lipsum}
\usepackage{amsfonts}
\usepackage{amssymb}
\usepackage{graphicx}
\usepackage{xcolor}
\journal{Optics Communications}

\begin{document}

\begin{frontmatter}

\title{ Change of polarization degree of light beams on propagation in curved space}
\author[1]{You-Lin Chuang\corref{cor}}
\ead{chuang_youlin@gapp.nthu.edu.tw}
\author[1]{Himanshu Parihar}
\cortext[cor]{Corresponding author}
\affiliation[1]{organization={Physics Division, National Center for Theoretical Sciences},
city={Taipei},
postcode={10617},
country={Taiwan}}
%\date{\today}
\begin{abstract}
Even in free space, which is commonly considered of as a flat space-time in most settings, the degree of polarization of a partially spatially coherent light beam changes as it travels. 
Similarly, the polarization degree would change when a partially spatially coherent light beam propagates in a curved space-time. 
The difference of the polarization degree between the curved space and flat space can reveal the essential structure of the curved space.
In this work, we consider a simplest case of curved space known as Schwarzschild spacetime.  
We can simulate the Schwarzschild space-time as an optical material with an effective refractive index. The difference of the polarization degree of a light beam propagating in curved space and flat space can be achieved up to $ 5\% $, which is detectable in practical measurement. In addition, we have found that the partially spatially coherent light source is necessary for obtaining significant changes in polarization degree.  
Our results provide an alternative method to estimate the Schwarzschild radius of a massive object with the optical polarization degree measurement. 
\end{abstract}

%\begin{graphicalabstract}
%\end{graphicalabstract}


\end{frontmatter}

%%%%%%%%%%%%%%%%%%%%%%%%%%%%%%%%%%%%%%%%%%%%%%%%%%%%%%%%%%%%%%%%%%%%%%%%%%%%%%%%%%%%%%

\section{Introduction}

It is well known that the correlation of a light field at any two points may changes drastically as it propagates through free space. This is so called van Cittert-Zernike theorem \cite{vanCittertheorem}. The most extreme case is that, the spatial correlation on the observation plane will be non-zero even the light source is spatially incoherent.     
As the correlation between two points, the spectrum of the light, which is represented the diagonal elements in the cross-spectral density has a (red- or blue-) shift in the far zone due to the correlation, and this effect of ``correlation induced spectral changes" is known as the Wolf effect or Wolf shift, which was first predicted by Emil Wolf in 1987, and was confirmed in experimental observation soon after the theoretical prediction \cite{invariance, redshift}.  
Similar than the Wolf effect, it has been known that the polarization state of a light beam with partially coherence in its transverse plane will changes even when it propagates through free space \cite{flat}. From the cross-spectral density matrix, one can find an intimate relation between the degree of coherence and the degree of polarization, which implies that the degree of polarization of a partially coherent light beam on the observation plane directly depends on the degree of coherence and polarization on the source plane \cite{correlation}. The discussions of the polarization degree changes in the cases of Young's interference experiment \cite{Young1, Young2}, the diffraction from a circular aperture \cite{Fraunhofer, exp}, and general paraxial optical systems \cite{opticalsystem} have been studied in the past decade.     

In recent decades, some pioneer works for optics in curved space had been studied.
\cite{curvedspaceoptics1, curvedspaceoptics2}. The discussions of light propagation in curved surface had been considered\cite{curvedsurface1, curvedsurface2, curvedsurface3}. 
In quantum optics, the intensity correlation proposed by Hanbury Brown and Twiss (HBT) has revealed that the behaviors of spatial coherence in curved space are quite different than that of in flat space \cite{HBTcurve}, and also been proposed for the measurement of space curvature.
With the essential difference of the space structure for the different curvatures, one can expect that the optical correlations would have different behaviors when the light beam is travelling through it. Wolf effect of partially coherent light fields in two-dimensional curved space was studied in theory \cite{wolfcurve}, and showed that the space curvature has strong influence on the light spectra during the light propagations in curved space. 

Even though there are so many research works on optics in curved space, people restrict their studies on curved surface, which is a two-dimensional space embedded in three-dimensional space. However, some fundamental optical problems such as the optical polarization and light propagations are generally considered in three-dimensional space, especially for the polarization, which becomes more complicated for a light confined on surface.     
On the other hand, from the general relativity it is well known that the space-time is essentially curved, especially near a massive object, and has been proved by many observations in astrophysics \cite{blackhole, gravitationalwave}.   
Therefore, it is natural to consider the optical problems in the genuine curved three-dimensional space near a massive object such as black holes or neutron stars.  
In this note, we are motivated by the intimate connections between coherence and polarization. We consider the polarization degree of a partially coherent light beam in curved space arisen from a black hole, studying how the curved space affects the polarization degree of light beam when it propagates through it.

\section{Theory}

The Einstein's theory of general relativity describes gravity in terms of the geometry of space-time in the presence of mass and energy. A free particle in the absence of a gravitation field moves in shortest path usually a line in Euclidean space. However, in the presence of gravitational field, particles also move in a shortest path known as geodesic which is the generalization of a straight line in a curved space. The Einstein's field equation in general relativity relates the curvature of space-time to the matter distribution present in it. There are many known solutions of Einstein field equation of which the Schwarzschild solution is the simplest solution that describes the gravitational field in spherically symmetric space (outside the spherical body). The Schwarzschild metric is given by 
\begin{equation}
ds^2=-\left(1-\frac{2GM}{c^2r}\right)c^2dt^{2}+\left(1-\frac{2GM}{rc^2}\right)^{-1}dr^2
+r^{2}d\Omega^{2},
\end{equation}
where $M$ is the mass of the black hole, $G$ is the gravitational constant, $c$ is the speed of light and $d\Omega$ is the metric on the sphere.
It is a static, stationary and spherically symmetric solution. The Schwarzschild radius is given by $r_s=2GM/c^2$ and the location $r=r_s$ is known as the event horizon of the Schwarzschild black hole. The nature of the space-time inside the event horizon is different from the region outside the event horizon. Any gravitating body whose radius is less than or equal to its Schwarzschild radius, is called a black hole. For an asymptotic observer all the events outside $r=r_s$ freeze at the event horizon.

\begin{figure}[t]
\begin{center}
\includegraphics[scale=0.7]{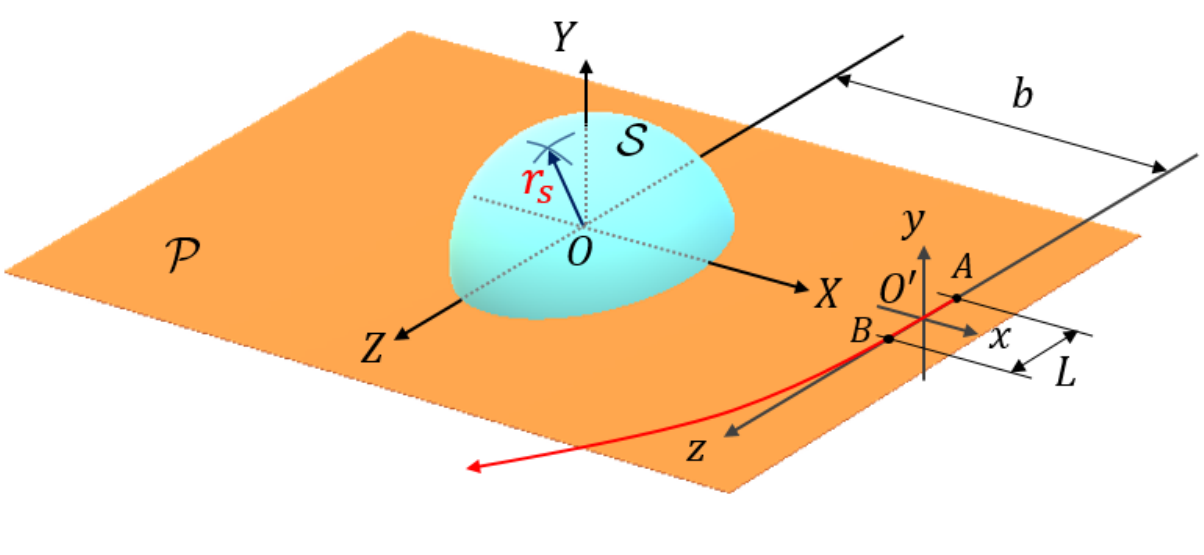} 
\end{center}
\caption{The light beam passing through a Schwarzschild black hole. }
\label{curvespacetime}
\end{figure}

Let's consider the problem of light field propagation in a curved space-time due to the black hole. As shown in Fig. \ref{curvespacetime}, a black hole whose center locates at point $ O $, and the center of the coordinate system of the light beam is at point $ O' $. 
According to the theory of general relativity, the massive object can curve the space and time in the vicinity so that the trajectory of light beam is deflected toward the massive object when the light beam is passing by. This is the main physics behind the phenomenon of gravitational lensing effect. 
The Maxwell's equations in curved space-time are very complicated and difficult to find solutions with an intuitive physical picture.    
Therefore, here we map the curved space into an effective optical problem instead. We treat the curved space as an optical material with an effective refractive index $ n(r) $ \cite{effectiven1, effectiven2, effectiven3}, whose explicit form is given by
\begin{eqnarray}
n(r) = \left( 1-\dfrac{r_s}{r}\right)^{-1}, 
\label{n}
\end{eqnarray}
where $ r $ is the distance between any point in the space and the origin $ O $.
As we can see from Eq. (\ref{n}), $ r $ must be greater than $ r_s $, which is the bolder of the event horizon.   

In our case, we arrange the light source at point $ A $, and the detector at point $ B $ for observation. The distance between them is $ L $, which is very short compared to the Schwarzschild radius $ r_s $ so that the deflection of light can be negligible. Besides, we can always find a plane $ \mathcal{P} $ which contains the trajectory of light beam and the great circle of the sphere $ \mathcal{S} $.  
In this scenario, we can define the two coordinate systems, which are $ (X,Y,Z) $ and $ (x,y,z) $, whose origins are located at $ O $ and $ O' $, respectively. 
$ O $ and $ O' $ are also on the same plane $ \mathcal{P} $, for which we set $ y =Y = 0 $. As one can see in Fig. \ref{curvespacetime}, the transformation of the two coordinate systems is given by $x = X-b $ and $ z = Z $. 
  
With the transformation, we can write down the effective refractive index in the light beam coordinate. That is 
\begin{eqnarray}
 n = n(x,z) = \left[ 1 - \dfrac{r_s}{\sqrt{\left( x+b\right)^2 +z^2}}\right]^{-1}. 
\label{n_light}
\end{eqnarray} 
Since the scale of $ x $ is about the diameter of the light beam, which is much smaller than the distance of $ b $. Thus we can approximate Eq. (\ref{n_light}) by ignoring the transverse variables. Thus we can obtain
\begin{eqnarray}
n \simeq n(z) = \left[ 1 - \dfrac{r_s}{\sqrt{b^2+z^2}}\right]^{-1}.
\label{n_approxi}
\end{eqnarray}
  
From Eq. (\ref{n_approxi}), we can envision that the light beam is propagating through an optical medium with the refractive index varying in the longitudinal direction.  
Because of the absence of the transverse variables in the effective refractive index , we can simulate the system as a series of parallel plates for each one is at $ z_i $ and thinckness is $ \Delta z_{i} $, and has the refractive index $ n_i = n(z_i) $. 
We can calculate the ABCD matrix in this system by taking limits of $ N\rightarrow\infty $ and $ \Delta z_{i}\rightarrow dz $, where $ N $ is the number of plates. Then we have found that 
\begin{eqnarray}
M = \left( \begin{array}{cc}
A & B\\
C & D
\end{array} \right) 
= \left( \begin{array}{cc}
1 & B(z)\\
0 & 1
\end{array} \right), 
\end{eqnarray} 
where 
\begin{eqnarray}
B(z) = \int_{-l}^z \dfrac{dz'}{n(z')} = (z+l) + r_s \ln\left( \dfrac{-l+\sqrt{b^2+l^2}}{z+\sqrt{b^2+z^2}}\right). 
\label{B}
\end{eqnarray}
Here we consider a light beam is emitted from point $ A $, which is at $ z = -l $, and one can observe the light beam at point $ B $, which is at $ z' = +l $, and $ l = L/2 $.
The propagation of an electromagnetic field passing through an ABCD optical system is described by the Huygens-Collins diffraction integral \cite{ABCD1, ABCD2}.
\begin{eqnarray}
E_i(x,y,z) = -\dfrac{ik}{2\pi B(z)}\iint_{-\infty}^{+\infty}dx'dy'E_i(x', y',0)\exp\left\lbrace \dfrac{ik}{2B(z)}\left[ (x-x')^2+(y-y')^2\right]  \right\rbrace,
\label{Ef}    
\end{eqnarray}   
in which we have used $ A = 1 = D $ and $ C = 0 $. $ i\in{x,y} $ belongs to the two field  polarizations. The integral operates on $ x' $ and $ y' $, which is on the incident plane, i.e. $ z = 0 $.

As we can see in Eq. (\ref{Ef}), the only one difference from the case of field propagation in the empty (flat) space is the replacement of $ z $ by the factor $ B(z) $.  
The degree of coherence and the polarization of the partially spatially coherent field would be directly affected by the factor $ B(z) $.  
Based to Eq. (\ref{Ef}), one can construct the vectorial cross-spectral density matrix at $ z>0 $.  
\begin{eqnarray}
&&\textbf{W}(\bm{\rho}_1,\bm{\rho}_2;z)\equiv W_{ij}(\bm{\rho}_1,\bm{\rho}_2;z) = \left\langle E_i^{\ast}(\bm{\rho}_1,z)E_j(\bm{\rho}_2,z)\right\rangle \nonumber\\
&&= \iint d^2\rho_1'\iint d^2\rho_2'~h^{\ast}(\bm{\rho_1},\bm{\rho_1'};z)h(\bm{\rho_2},\bm{\rho_2'};z)W^{(0)}_{ij}\left( \bm{\rho_1'}, \bm{\rho_2'} \right), \nonumber\\   
\label{Wij}
\end{eqnarray}
where $ \bm{\rho}_i \equiv (x_i,y_i) $ and $ \bm{\rho}_i' \equiv (x_i',y_i') $, $ i\in{1,2} $ are the two points on the transverse plane. $ W^{(0)}_{ij} $ is the cross-spectral density for the incident field, and the impulse response function is given by 
\begin{eqnarray}
h(\bm{\rho},\bm{\rho'};z) = \left[-\dfrac{ik}{2\pi B(z)}\right] \exp\left[ \dfrac{ik}{2B(z)}\left( \bm{\rho}-\bm{\rho'}\right)^2\right]. 
\label{h}
\end{eqnarray}
The degree of coherence $ \eta $ and polarization $ P $ \cite{polarization} can be directly calculated from Eq. (\ref{Wij}). That is
\begin{eqnarray}
&&\eta(\bm{\rho}_1,\bm{\rho}_2;z) = \dfrac{{\rm{Tr}}\textbf{W}(\bm{\rho}_1,\bm{\rho}_2;z)}{\sqrt{{\rm{Tr}}\textbf{W}(\bm{\rho}_1,\bm{\rho}_1;z)}\sqrt{{\rm{Tr}} \textbf{W}(\bm{\rho}_2,\bm{\rho}_2;z)}}, \label{doc}\\
&& P(\bm{\rho};z) = \sqrt{1-\dfrac{4 {\rm{Det}}\textbf{W}(\bm{\rho},\bm{\rho};z)}{\left[ {\rm{Tr}} \textbf{W}(\bm{\rho},\bm{\rho};z)\right]^2 }}. \label{P}
\end{eqnarray} 

$ W^{(0)}_{ij}\left( \bm{\rho_1'}, \bm{\rho_2'}\right) $ in Eq. (\ref{Wij}) is the cross-spectral density matrix on the source plane.
Once the statistical averages of the partially coherent light beam are given at $ z=0 $, one can determine that of light beam propagating to the observation plane. 
Here we consider that the partially coherent light beams on the source plane can be described by Gaussian-Schell model (GSM) \cite{GSM1, GSM2}, and the cross-spectral density can be expressed as 
\begin{eqnarray}
W^{(0)}_{ij}\left( \bm{\rho_1'}, \bm{\rho_2'}\right) = \sqrt{S_i(\bm{\rho_1'})}\sqrt{S_j(\bm{\rho_2'})}\mu_{ij}\left( \bm{\rho_1'}, \bm{\rho_2'}\right), \label{W0}
\end{eqnarray}
in which
\begin{eqnarray}
&&S_i(\bm{\rho}) = s_i(\omega)\exp\left( -\bm{\rho}^2/2\sigma_i^2 \right), \label{Si}\\
&&\mu_{ij}\left( \bm{\rho_1}, \bm{\rho_2}\right) = B_{ij} \exp\left[ -(\bm{\rho}_1-\bm{\rho}_2)^2/2\delta_{ij}^2 \right],   \label{muij}
\end{eqnarray}
where $ s_i $ and $ \sigma_i $ are the magnitude and beam width of the field component. $ \delta_{ij} $ describes the coherent length of light beam. $ B_{ij} = 0 $ as $ i\neq j $, and $ B_{ij} = 1 $ as $ i=j $. $ \mu_{ij} $ denotes the coherence degree of the light beam. 
For simply, we consider $ \sigma_x = \sigma_y = \sigma $. The coherence length in $ x $ and $ y $ directions are $ \delta_{xx} \equiv \delta_x $ and $ \delta_{yy}\equiv\delta_y $. $ \delta_{xy} = 0 = \delta_{yx}$.
In addition, we introduce a factor $ \alpha \equiv s_y/s_x $ to quantify the ratio between the field strengths of $ x $ and $ y $  
components. From Eq. (\ref{W0}) with Eq. (\ref{Si}) and Eq. (\ref{muij}), 
the matrix form of cross-spectral density of the light beam on the source plane is shown as
\begin{eqnarray}
&&W^{(0)}_{ij}\left( \bm{\rho_1'}, \bm{\rho_2'}\right) = s(\omega)\exp\left[ -\left( \bm{\rho_1'}^2+\bm{\rho_2'}^2\right)/4\sigma^2 \right] \nonumber\\
&&\times \left[ 
\begin{array}{cc}
\exp\left[ -\dfrac{(\bm{\rho_1'}-\bm{\rho_2'})^2}{2\delta_x^2}\right] & 0 \\
0 & \alpha\exp\left[ -\dfrac{(\bm{\rho_1'}-\bm{\rho_2'})^2}{2\delta_y^2}\right]
\end{array}\right], \label{matrixform} 
\end{eqnarray} 
where we have set $ s_x(\omega) \equiv s(\omega) $. From the expression given in Eq. (\ref{P}), we can obtain the degree of polarization on the source plane.
\begin{eqnarray}
P^{(0)}(\bm{\rho}) = \left\vert \dfrac{1-\alpha}{1+\alpha}\right\vert .
\label{P0}
\end{eqnarray} 
According to Eq. (\ref{P0}), it shows clearly that when $ \alpha = 1 $ there is no any polarization degree on the source plane.

We are interested in the problem of how the polarization degree of the field changes after propagating through the curved space due to the black hole. In order to obtain the cross-spectral density matrix at $ z' $, we substitute Eq. (\ref{matrixform}) into Eq. (\ref{Wij}), and we can obtain
\begin{eqnarray}
&&W_{xx}\left( \bm{\rho}, \bm{\rho}; z\right) = s(\omega)\left( \dfrac{1}{\beta_x^2(z)}\right)  \exp\left( -\dfrac{\bm{\rho}^2}{2\sigma^2\beta_x^2(z)}\right),  \nonumber\\ 
&&W_{yy}\left( \bm{\rho}, \bm{\rho}; z\right) = \alpha s(\omega)\left( \dfrac{1}{\beta_y^2(z)}\right) \exp\left( -\dfrac{\bm{\rho}^2}{2\sigma^2\beta_y^2(z)}\right), \nonumber \\
&&W_{xy}\left( \bm{\rho}, \bm{\rho}; z\right) = 0 = W_{yx}\left( \bm{\rho}, \bm{\rho}; z\right), \nonumber\\ \label{Wz}
\end{eqnarray}
and
\begin{eqnarray}
\beta_i^2(z) = 1 + \dfrac{B^2(z)}{4k^2\sigma^4} + \dfrac{B^2(z)}{k^2\sigma^2\delta_i^2},~~~i\in {x,y}. \label{beta}
\end{eqnarray}
From Eq. (\ref{Wz}), the degree of polarization of the light beam is 
\begin{eqnarray}
P(\bm{\rho},z) = \left\vert \dfrac{W_{xx}\left( \bm{\rho}, \bm{\rho}; z\right) - W_{yy}\left( \bm{\rho}, \bm{\rho}; z\right)}{W_{xx}\left( \bm{\rho}, \bm{\rho}; z\right) + W_{yy}\left( \bm{\rho}, \bm{\rho}; z\right)}\right\vert . \label{Pz}
\end{eqnarray}
When the black hole is absent, i.e. $ r_s = 0 $, we have $ B(z) = z $, and our results reduce to the results in flat space \cite{flat}.     
According to Eq. (\ref{Pz}) with the helps of Eq. (\ref{Wz}) and (\ref{beta}), we are in the position to study how the Schwarzschild black hole influences the degree of polarization of a light beam. 

\section{Results and Discussions}

In last section, we have used the effective refractive index approach to simulate the curved space arisen from the Schwarzschild black hole. With the fact of the small length scale in the transverse of light beam, we can obtain the resulting refractive index given in Eq. (\ref{n_approxi}), and then the factor $ B(z) $ can also be calculated by Eq. (\ref{B}). This is the model for us to study how the polarization degree of the partially spatially coherent light changes in the curved space. 

\begin{figure}[t]
\begin{center}
\includegraphics[scale=0.455]{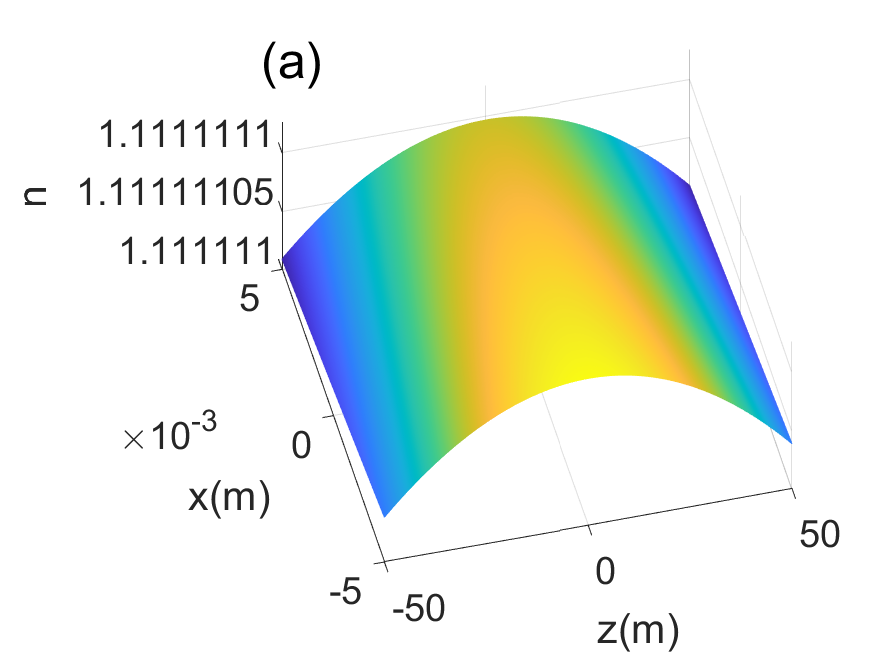} 
\includegraphics[scale=0.455]{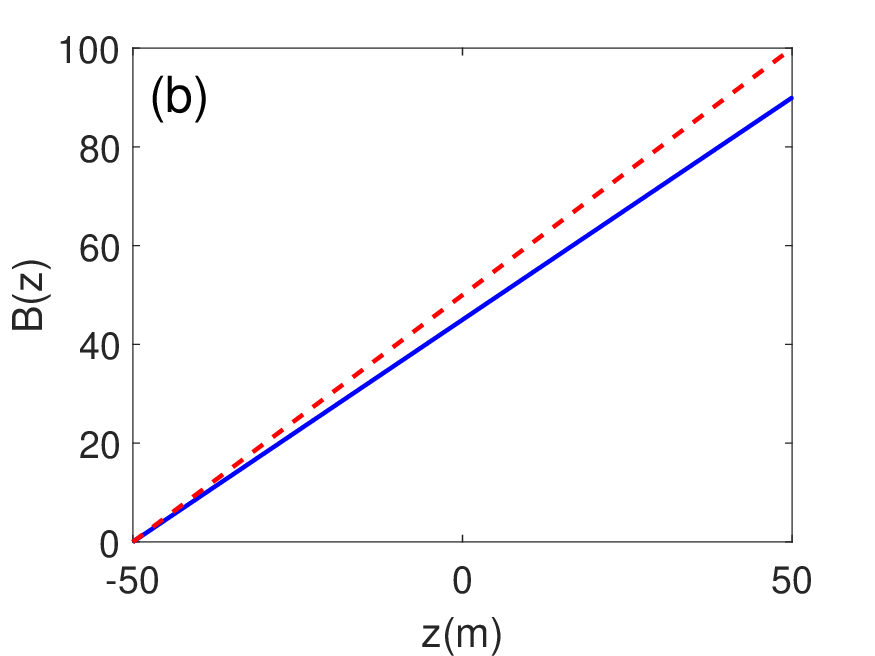} 
\end{center}
\caption{ (a) The refractive index $ n(x,z) $. (b) $ B(z) $ versus $ z $. The blue line corresponds to the case with the curved space arisen from the black hole, and red dashed line represents the case of light propagating through the flat space. For the two plots, we use $ b = 10r_s $, and $ r_s = 4,200~\rm m $. } 
\label{nB_plot}
\end{figure}

In Fig. \ref{nB_plot}, we plot the two functions, which are given in Eq. (\ref{n_light}) and (\ref{B}). In Fig. \ref{nB_plot}(a), we plot the distribution of the effective refractive index. The range in $ x $ direction is from $ -5\sigma $ to $ +5\sigma $, where $ \sigma = 1~\rm mm $ is the beam width of light beam. The range of $ z $ direction is from $ -50~\rm m $ to $ +50~\rm m $. As we can see, the refractive index is almost the same in $ x $ direction, and the main change in refractive index is along the propagation direction $ z $. 
In Fig. \ref{nB_plot}(b), we plot the function of $ B(z) $ versus $ z $. The blue line corresponds to the case with the curved space arisen from the black hole, and red dashed line represents the case of light propagating through the flat space. 
It is obvious that the blue line deviates from the red dashed line. The small change in $ B $ will reflect in the polarization degree of light beam. The parameters we use here are $ r_s = 4,200~\rm m $, which is the typical Schwarzschild radius of a neutron star, and $ b = 10r_s $. 

\begin{figure}
\begin{center}
\includegraphics[scale=0.7]{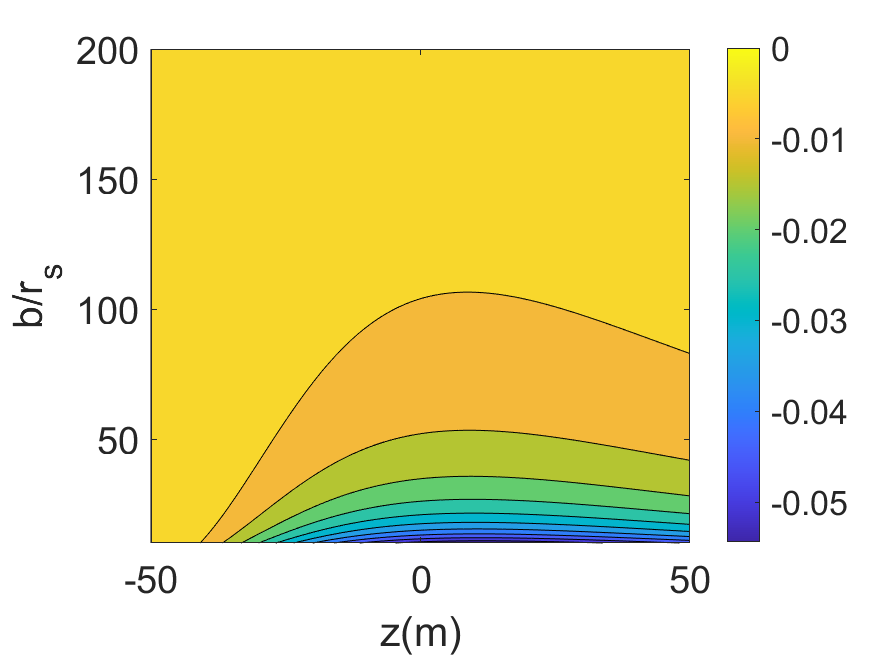} 
\end{center}
\caption{The contour plot of the polarization degree change versus $ z $ and $ b $. The field parameters are used by $ \lambda = 10^{-12} \rm m $, $ \delta_x = 10^{-8} \rm m $, $ \delta_y = 10\delta_x = 10^{-7} \rm m $, and $ \alpha = 2 $. }
\label{contour}
\end{figure}

Next, we would like to study how the field polarization degree changes in the curved space. 
In order to have a significant change in polarization degree, we consider the light beam in high frequency. We use $ \lambda = 10^{-12}\rm m $, which is in $ \gamma $-ray region.   
The coherence lengths in $ x $ and $ y $ directions are $ \delta_x = 10^{-8} \rm m $ and $ \delta_y = 10\delta_x = 10^{-7} \rm m $. $ \alpha = 2 $. These parameters satisfy the condition of Gaussian Schell-model beams. 
In Fig. \ref{contour}, we show the contour plot for the percentage of polarization degree change with respect to $ z $ and $ b $. Here we define the percentage of polarization degree change as 
$ \xi\equiv(P-P_{\rm flat})/P_{\rm flat} $, in which $ P_{\rm flat} $ is the field polarization degree in flat space. We set $ P_{\rm flat} $ as a reference, and compare the polarization degree in curved space with respect to that in flat space. Initially,
we can find that $ \xi $ is close to zero when $ z = -50~\rm m $ where is the position of light source. 
The polarization degree change becomes non-zero when $ z $ is close to the nearest point.  
When $ b $ is smaller, we can observe larger polarization degree change. It can achieve about $ 5\% $ polarization degree change, which is possible to detect in practical observation.  
The polarization degree change gradually decreases when $ b $ is getting larger and larger. 

\begin{figure}
\begin{center}
\includegraphics[scale=0.455]{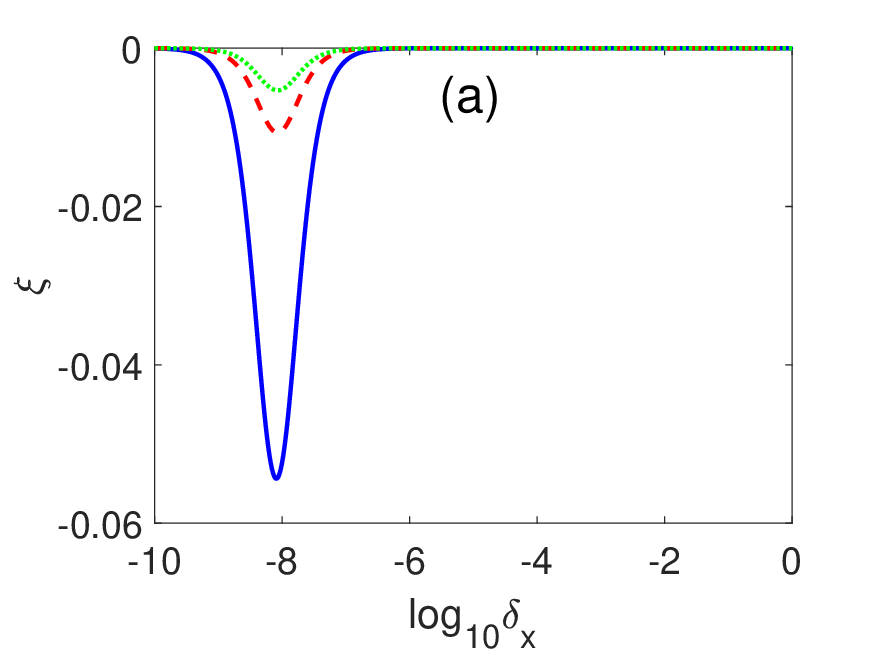} 
\includegraphics[scale=0.455]{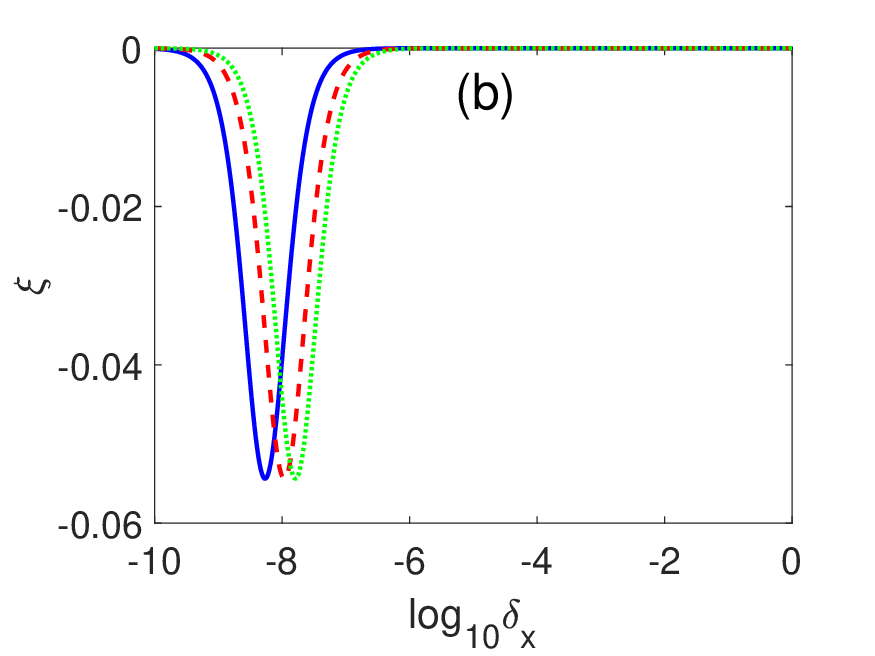} 
\end{center}
\caption{(a) $ \xi $ v.s. $ \log_{10}(\delta_x) $ at $ z = 0 $. 
The blue, red dashed, and green dotted curves are plotted with $ b/r_s = 10,$ 50, and 100, respectively. (b) $ \xi $ v.s. $ \log_{10}(\delta_x) $ at $ b/r_s = 10 $. The red dashed, blue, and green dotted curves are the results at $ z/L=  $ 1/3, 2/3, 1, respectively.}
\label{dxbz}
\end{figure}

The coherence length $ \delta $ is an other important parameter in the partially spatially coherent light. There are two extreme cases. The full coherence is the extreme case when the coherence length goes to infinite, and when $ \delta\rightarrow 0 $ it becomes incoherent light.  
In order to know how the coherence length effect on the polarization degree change, we plot the value of $ \xi $ with respect to $ \delta_x $ under different values of $ b $ at $ z=0 $ in Fig. \ref{dxbz}(a). The three curves, which are blue, red dashed, and green dotted curves, correspond to $ b/r_s = $10, 50, and 100, respectively. The ratio of $ \delta_y/\delta_x = 10 $ is kept. 
As we can see, we have a range of coherence degree $ \delta_x $ to get significant polarization degree change. In our case, the range is about $ 10^{-7.5} $ to $ 10^{-8.5} $. 
Besides, $ \xi $ is getting larger when $ b $ is smaller. 
It is clear to understand that the stronger gravity due to the massive object can make larger polarization degree changes.
In Fig. \ref{dxbz}(b), we plot $ \xi $ versus $ \delta_x $ at $ b = 10r_s $ under different $ z $'s. The red dashed, blue, and green dotted curves are the results at $ z/L=  $ 1/3, 2/3, 1, respectively. The coherence $ \delta_x $ for the maximum of $ \xi $ shifts a little to higher coherence. 
From the results shown in Fig. \ref{dxbz}, we have an important conclusion that the spatially incoherent light plays a crucial role to obtain a significant polarization change.   
 
\section{Conclusion}

We have studied the polarization degree change of a partially spatially coherent light propagating through a curved space with Schwarzschild space-time. 
With the approach of effective refractive index, we can include the spatial coherence information which is related to the polarization degree when the light beam propagates through the curved space.    
The polarization degree change can be achieved upon to $ 5\% $, which is detectable in practical measurement. In addition, we also point out that the spatial partially coherence is an indispensable resource for obtaining the significant changes in polarization degree. 
The scheme of observing the polarization change with the partially spatially coherent light provides an optical method to infer the Schwarzschild radius of a massive object.

\section*{ Acknowledgement }

The authors would like to thank National Tsing Hua university for MATLAB software license and National Science and Technology Council for financial support.

%%%%%%%%%%%%%%%%%%%%%%%%%%%%%%%%%%%%%%%%%%%%%%%%%%%%%%%%%%%%%%%%%%%%%%%%%%%%%%%%%%%%%%%%%

\end{document}